\documentclass[structabstract]{aa}
\usepackage{graphicx,amsmath,natbib,txfonts}
\usepackage{natbib}
\usepackage[dvips]{changebar}

\newcommand\lta{\mathrel{\hbox{\raise 0.6 ex \hbox{$<$}\kern
                   -1.8 ex\lower .5 ex\hbox{$\sim$}}}}
\newcommand\gta{\mathrel{\hbox{\raise 0.6 ex \hbox{$>$}\kern
                   -1.7 ex\lower .5 ex\hbox{$\sim$}}}}
 
\newcommand{\scrbox}[1]{\ensuremath{{\mbox{\scriptsize #1}}}}

\newcommand{\teff}{{\ensuremath{T_{\scrbox{eff}}}}}
\newcommand{\Msol}{\ensuremath{\,\mbox{\rm M}_{\odot}}}
\newcommand{\Mloss}{\ensuremath{\,\mbox{\rm M}_{\odot}{\rm yr}^{-1}}}

\newcommand{\MS}{main--sequence}
\newcommand{\gr}{\ensuremath{g_{\scrbox{rad}}}}

\newcommand{\DM}{\ensuremath{ \log \Delta M/M_{*}}}
\newcommand{\Dm}{\ensuremath{\Delta M/M_{*}}}

\begin{document}

\title{Abundance anomalies in pre--main-sequence stars}
\subtitle{Stellar evolution models with mass loss}

\author{M. Vick\inst{1,}\inst{3}, G. Michaud\inst{2,}\inst{3}, J. Richer\inst{3}, 
 \and O. Richard\inst{1}}

\institute{GRAAL UMR5024, Universit\'e Montpellier II,
                 CC072, Place E. Bataillon,
                 34095\,Montpellier Cedex 05,
                 France
\and
            LUTH, Observatoire de Paris,
	    CNRS, Universit\'e Paris Diderot,
	    5 Place Jules Janssen, 92190\,Meudon, France           
\and
	   D\'epartement de physique, Universit\'e de Montr\'eal, 
           Montr\'eal, Qu\'ebec, H3C 3J7, Canada\\
           \email{mathieu.vick@umontreal.ca, michaudg@astro.umontreal.ca,\\
                  jacques.richer@umontreal.ca, olivier.richard@graal.univ-montp2.fr}
}

\date{Received August 5th, 2010; Accepted October 10th, 2010}

\abstract{}
{The effects of atomic diffusion  
on internal and surface abundances of A and F
pre--main-sequence stars with mass loss are studied in order to determine at 
what age the effects materialize, as well as to further understand 
the processes at
play in HAeBe and young ApBp stars.}
{Self-consistent stellar evolution models of 1.5 to 2.8\Msol{}
with atomic diffusion (including radiative accelerations) for all species 
within the OPAL opacity database were computed and compared to observations of
HAeBe stars.}
{Atomic diffusion in the presence of weak mass loss 
can explain the observed abundance anomalies of pre--main-sequence stars, as well as the
presence of binary systems with metal rich primaries and chemically normal
secondaries such as V380 Ori and HD72106. This is in contrast to 
turbulence models which do not allow for abundance anomalies to develop on the
pre--main-sequence. The age at which anomalies can appear depends on stellar mass.}
{For A and F stars, the effects of atomic diffusion can modify both the internal and 
surface abundances before the onset of the \MS. The appearance of important surface abundance anomalies
on the pre--main-sequence does not require mass loss, though the mass loss rate affects their
amplitude. Observational tests are suggested to decipher 
the effects of mass loss from those of turbulent mixing. If abundance anomalies are confirmed
in pre--main-sequence stars they would severely limit the role of turbulence in these stars.}

\keywords{Diffusion --- stars: chemically peculiar --- stars: mass--loss --- stars:
pre--main-sequence --- stars: evolution --- stars: abundances} 

\maketitle

\section{Astrophysical context}
\label{sec:intro}
\begin{figure*}[!ht]
\begin{center}
\includegraphics[scale=1.0]{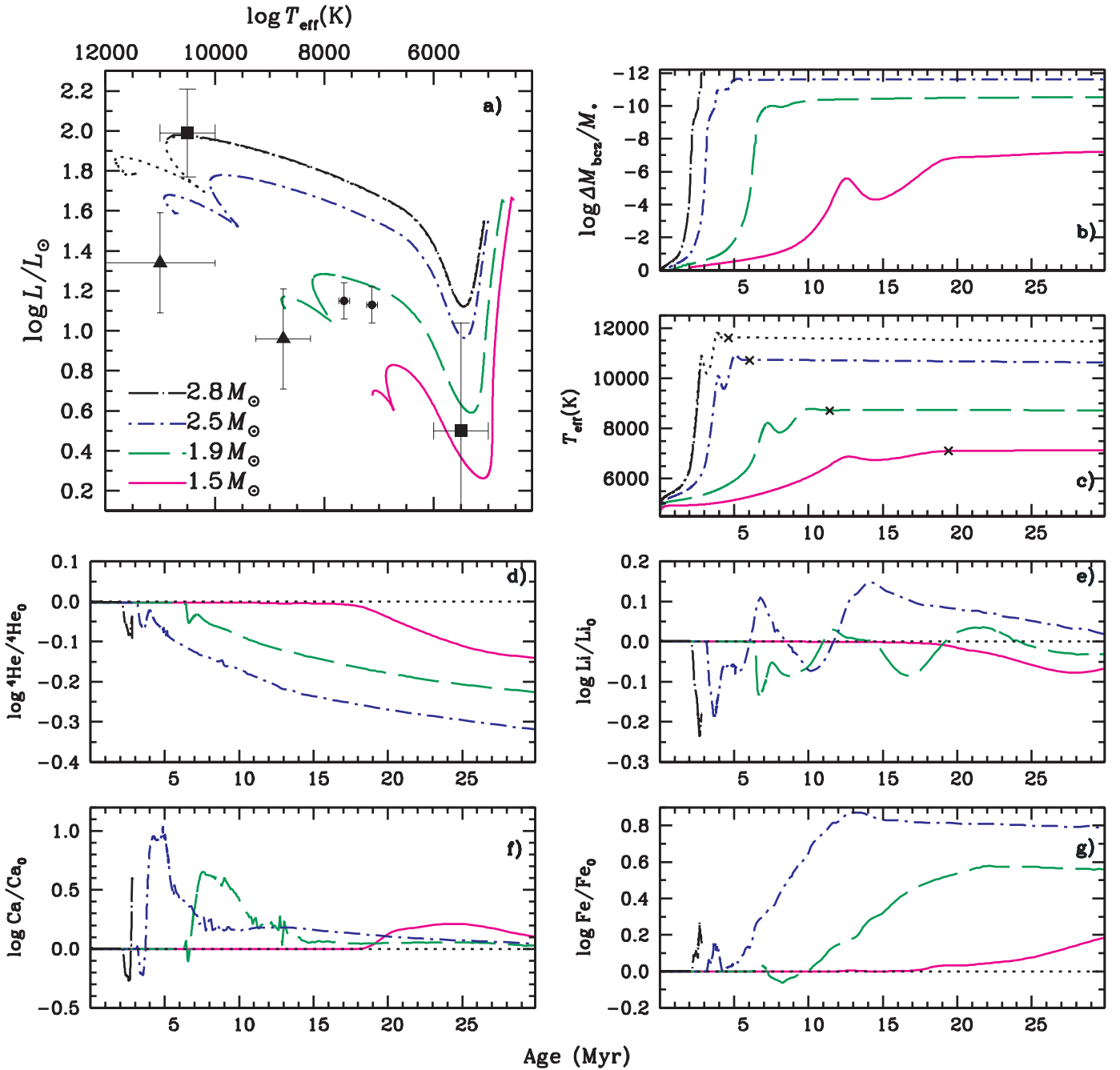}
\caption{The position in the HR diagram {\it (a)} is shown for 
four models with a mass loss rate of $5 \times 10^{-14}\Mloss$. The curves end (on the left) at
30 Myr, age at which all the models are on the \MS. The dotted segment 
of the 2.80\Msol{}
curve represents a model with a mass loss rate of $10^{-13}\Mloss$ 
and was added in {\it (a)} and {\it (c)} in order to facilitate extrapolation, 
though it was not added in
other panels for which the mass loss rate has an effect.
Observations are shown in the HR diagram for three sets
of binary stars: ($\blacksquare$) V380 Ori (\citealt{alecian09}); 
($\blacktriangle$) HD72106 (\citealt{folsom08}); 
($\bullet$) RS Cha (\citealt{alecian05}).
The evolution of the bottom of the
surface convection zone {\it (b)}, \teff\, {\it (c)}, as well as the 
abundances of He {\it (d)}, $^7$Li {\it (e)}, Ca {\it (f)} and 
Fe {\it (g)} are also shown. In panel {\it (c)}, ($\times$) marks the approximate end of the PMS.
}\label{fig:HR}
\end{center}
\end{figure*} 
The recent availability of magnetic field data 
from instruments such as ESPaDOnS at the 
Canada-France-Hawaii Telescope and Narval at the 
Bernard Lyot Telescope has allowed stellar physicists to
explore long standing questions in stellar physics. A flurry of 
recent studies have 
specifically focused on observing globally organized 
magnetic fields in intermediate mass pre--main-sequence (PMS) stars 
in order to  
determine the origin of magnetic fields in the chemically
peculiar Ap/Bp stars
\citep{donati97,alecian05,wade05,catala07,alecian08a,
alecian08b,folsom08,alecian09}. 
This is particularly interesting since A and B type stars 
have, at most, a very thin convective envelope, which is probably not 
sufficient to generate 
solar type, dynamo driven, magnetic fields. With this in mind, the
favored hypothesis used to explain the presence of important magnetic 
fields (from 100\,G to 10\,kG) in A and B \MS{} (MS) stars is the 
fossil field theory. 
In this context, the magnetic field would either have originated in the 
molecular cloud from which the star was born, or 
would have been generated by a 
dynamo process in the core during the star's earliest 
evolutionary stages. 
To test this hypothesis, the above mentioned 
studies have concentrated 
their efforts on 
charaterizing magnetic fields in Herbig (HAeBe) 
stars \citep{herbig60}, which 
are believed to be the PMS progenitors of Ap/Bp stars \citep{wade05}. 

In some cases, namely for HD72106, RS Cha and \hbox{V380 Ori}, 
binarity allows us to probe the effects of chemical separation since both
stars likely had the same initial composition.  
In fact, for at least two of these systems, the secondary has a solar metallicity, 
whereas the heavier primary star has an above solar metallicity and is likely
chemically anomalous. 
Since atomic diffusion timescales generally decrease when stellar mass increases, 
chemical separation could perhaps offer an explanation. 

Furthermore, a more thorough analysis may allow to characterize other phenomena
at play within these stars. What can the 
observed abundances tell us on the various processes, such as convection, 
mass loss and magnetic fields, which compete with atomic diffusion?  
And on what timescales can we expect significant surface abundance anomalies?
This paper will adress both of these questions.

In \citet{vick10} (hereafter Paper I),  
stellar evolution models with mass loss were 
introduced and shown to reproduce 
observed surface abundance anomalies for many AmFm stars.
However, observed abundance anomalies do not allow to determine
whether it is mass loss or turbulence which is competing with atomic diffusion 
within the radiative zone of these stars. It was nonetheless
established that the surface anomalies of AmFm stars were modulated by chemical
separation which occurs deep within the star.
Indeed, for both the mass loss models and
models with turbulent mixing
\citep[and references therein]{richer00,richard01}, chemical 
separation near $\DM\simeq -5$ to $-6$ is
responsible for the anomalous surface behavior. For PMS stars on the other hand,
chemical separation which occurs at this depth 
\hbox{cannot} explain observed abundance anomalies since the timescales are much too long.  
In mass loss models, chemical separation occuring near the surface
leads to anomalies which appear at the surface within a few Myr, 
and could therefore reconcile observations.    

In the following analysis, and in our calculations, 
mass loss is considered in {\it non rotating} stars, 
since in such stars, mass loss could be the only process competing with atomic 
diffusion within the stable radiative zones. An outline of the main aspects of our
evolution code is found in Sect.\,\ref{sec:calcul}. The stellar models will be 
presented in Sect.\,\ref{sec:models}, with particular 
attention to the various effects of mass loss and atomic diffusion 
in the interior (Sect.\,\ref{sec:structure}) and at 
the surface (Sect.\,\ref{sec:surface}) of PMS stars. 
In Sect.\,\ref{sec:observations},
the models are compared to observations of various HAeBe stars. Finally, in 
Sect.\,\ref{sec:conclusions}, the implications of our results on 
the processes involved in HAeBe and, by extension, in ApBp stars will be discussed.

\section{Calculations}
\label{sec:calcul}
The detailed description of the stellar evolution code used for these 
computations can be found in Paper I and references therein. 
At age zero, all models are fully convective
with the abundance mix prescribed in Table 1 of 
\citet{turcotte98soleil}. 
Opacities are continuously updated for every mesh point as abundances 
evolve. The only 
adjustable parameter, the mixing lenght parameter $\alpha$, was calibrated by 
fitting the current radius and luminosity of the Sun 
(see model H of \citealt{turcotte98soleil}). 
Radiative accelerations are taken from \citet{richer98} with corrections due 
to redistribution 
from \citet{gonzalez95} and \citet{leblanc00}. The introduction of mass 
loss and its impact on transport
are discussed extensively in Sect.\,4 of \hbox{Paper I}.

No extra mixing is enforced outside of convection zones. 
The effects of atomic diffusion materialize as soon as radiative 
zones appear as the models evolve toward the MS. These effects 
become more important as the radiative zone expands toward the surface, where 
atomic diffusion timescales are much shorter.

The unseparated\footnote{Having the same composition as the photosphere.} mass loss 
rates considered range from $10^{-14}$ 
to $10^{-13}$\Mloss: mass loss
rates which lead to surface abundances  
compatible with observations of many AmFm stars. 
Due to uncertainties related to the nature of 
winds for A and F stars (see discussion in Sect.\,4.2 of Paper I), we 
have chosen to limit our 
investigation to unseparated winds
in order to avoid introducing additional adjustable parameters.

Finally, this paper 
is part of a series of papers starting with \citet{turcotte98soleil}, 
where the 
mixing length used was calibrated 
using the Sun for given boundary conditions, helium and metal abundances.  
For consistency, and in order to isolate the effects of the processes of interest,
the same boundary condition, 
solar composition and mixing length are used for all calculations of Pop\,I stars (see also Sect.\,2 of Paper I).

\section{Evolutionary models}
\label{sec:models}

In Fig.\,\ref{fig:HR}, the position in the Hertzsprung--Russell (HR) diagram, the evolution of 
$\Delta M_{{\rm BSCZ}}$\footnote{BSCZ: bottom of surface convection zone.}, \teff\, 
as well as of the surface abundances of He, $^7$Li, Ca and Fe are shown for four
stellar models with the same mass loss rate ($5 \times 10^{-14}\Mloss$). 
Due to numerical
instabilities related to the complete disappearance 
of the surface convection zone, 
the 2.80W5E-14\footnote{The expression 2.80W5E-14 corresponds to a 2.80\Msol{} model 
with a mass loss rate
of $5 \times 10^{-14}\Mloss$.} model
could not be converged up to the \MS. Therefore, in order to complete 
the HR diagram, the
curve was continued using a 2.80W1E-13 model. In this case, 
the effect of doubling the mass loss rate  
on the position in the HR diagram 
is smaller than the width of the line. Its effect on 
surface abundances will be discussed in Sect.\,\ref{sec:surface}.

The exact definition of the MS and, by extension, the PMS is somewhat arbitrary. 
The definition given by  
\citet{iben65} stipulates that the \emph{zero age} \MS{} (ZAMS) begins when 
the thermal-gravitational energy is reduced to one percent of the 
luminosity of the star, although this may depend on such things as the treatment of convection 
\citep{cox68}. According to this definition, immediately after of the end of the
PMS, indicated by a ($\times$) in Fig.\,\ref{fig:HR}{\it c}, \teff{} and $L$ 
begin varying much more slowly. 
The time spent on the PMS varies from $\sim$4\,Myr for 
the \hbox{2.80W5E-14} model to $\sim$19\,Myr for the \hbox{1.50W5E-14} model.

\subsection{Radiative accelerations, internal abundances and structure}
\label{sec:structure}
\begin{figure}[!ht]
\begin{center}
\includegraphics[scale=0.45]{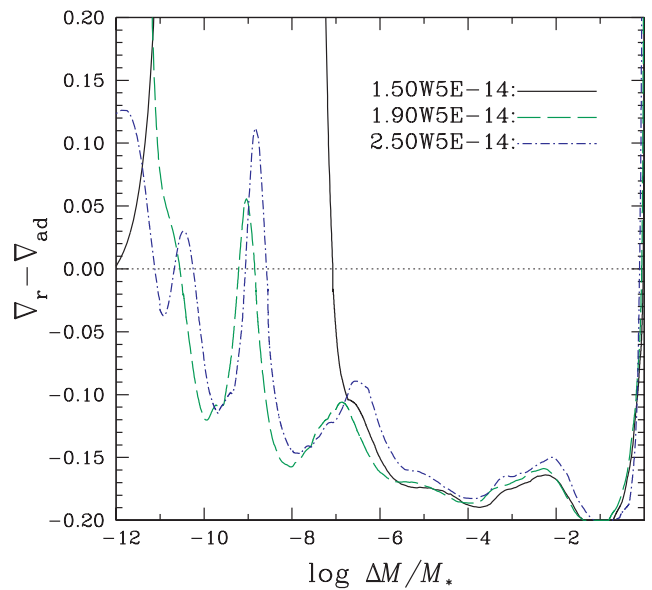}
\caption{Internal variation of $\nabla_{\rm r}-\nabla_{{\rm ad}}$ for 
three models with a mass loss rate
of $5 \times 10^{-14}\Mloss$ at the onset of the ZAMS. The surface is at \DM=$-12$ and 
transport is convective when $\nabla_{\rm r}-\nabla_{{\rm ad}}> 0$. For different masses, 
the position of convection zones in relation to \DM{} changes slightly, 
though it is constant in relation to $T$.
}\label{fig:grga}
\end{center}
\end{figure}
Even on the PMS, the internal structure and abundances vary significantly between 1.5\Msol{} and 2.8\Msol. 
In Fig.\,\ref{fig:grga}, convection zones are shown for models of different mass 
as they arrive on the ZAMS (see also Fig.\,\ref{fig:HR}{\it b}). The internal 
distribution of convection zones is
strongly correlated 
with stellar mass. For the 
1.50\Msol{} model, the SCZ includes 
the linked H and He convection zones and is never shallower than
\DM$=-7.2$ (see also Fig.\,\ref{fig:HR}{\it b}). 
For the 1.90W5E-14 model, 
the surface convection zones separate into a linked H$-$He\,I CZ and a deeper He\,II CZ, 
while a separate He\,I CZ materializes in
the heavier \hbox{2.50W5E-14} model. All convection zones disappear 
completely in the 2.80W5E-14 model 
before it has even arrived
on the MS (see Fig.\,\ref{fig:HR}{\it b}). Finally, over the entire PMS phase, 
the opacity bump near \DM$\sim -7.2$ resulting from iron peak element accumulation 
is not large enough to induce an iron peak convection zone for any of the three models, though 
iron peak opacity might extend the SCZ slightly inward for the 1.50W5E-14 model. If the mass loss rate
is $\leq 10^{-14}\Mloss{}$, an iron convection zone may appear after the PMS (see Fig.\,5 of Paper I). 

\begin{figure*}[!t]
\begin{center}
\includegraphics[scale=0.97]{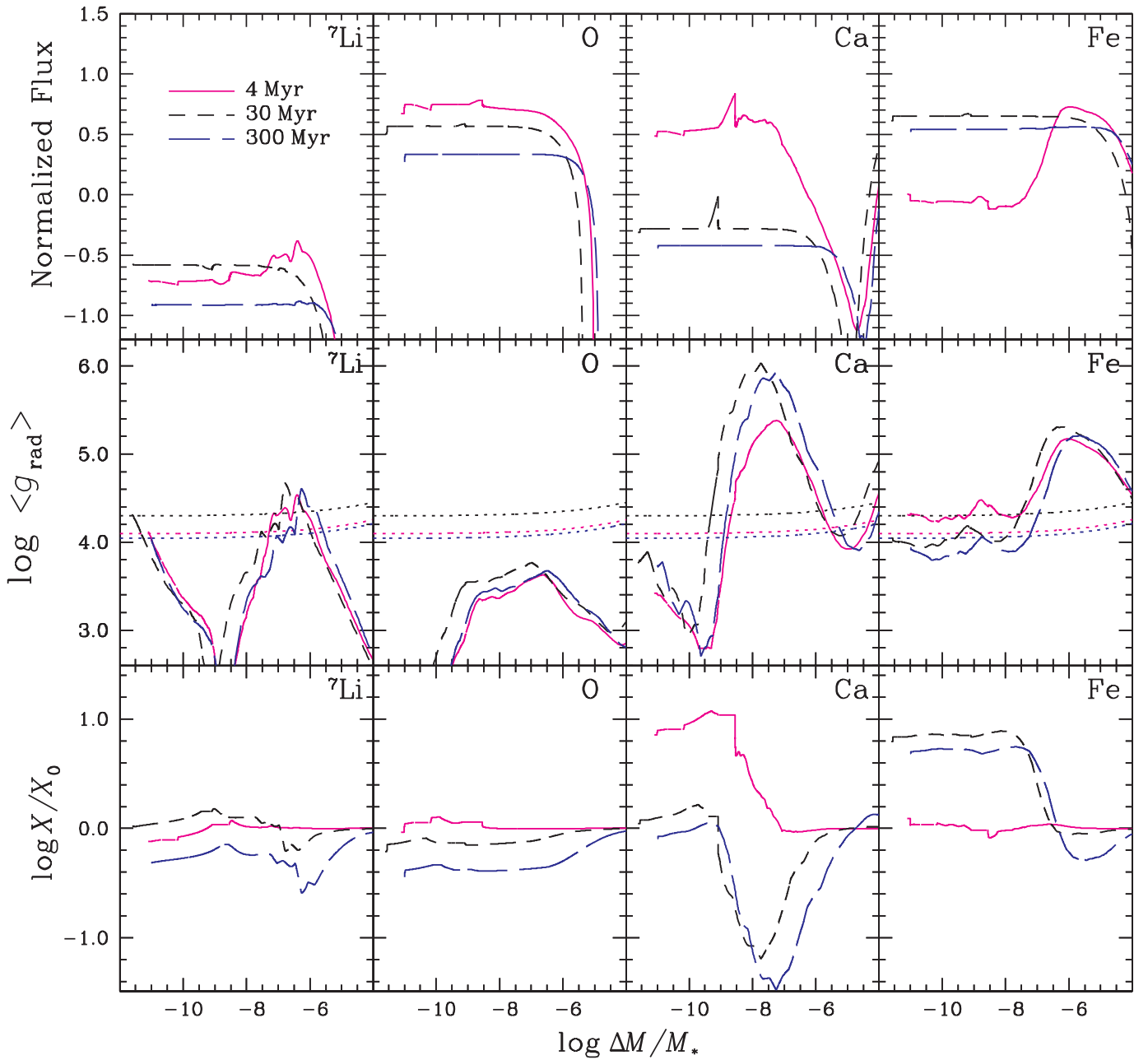}
\caption{Comparison of the normalized local flux with radiative accelerations and
internal abundances of $^7$Li, O, Ca and Fe at 3 different ages  
for the 2.50W5E-14 model. The curves end (on the left) at 
the bottom of the surface convection
zone. In the middle row, the dotted line represents gravity.}\label{fig:flux}
\end{center}
\end{figure*}
\begin{figure*}[!t]
\begin{center}
\includegraphics[scale=0.99]{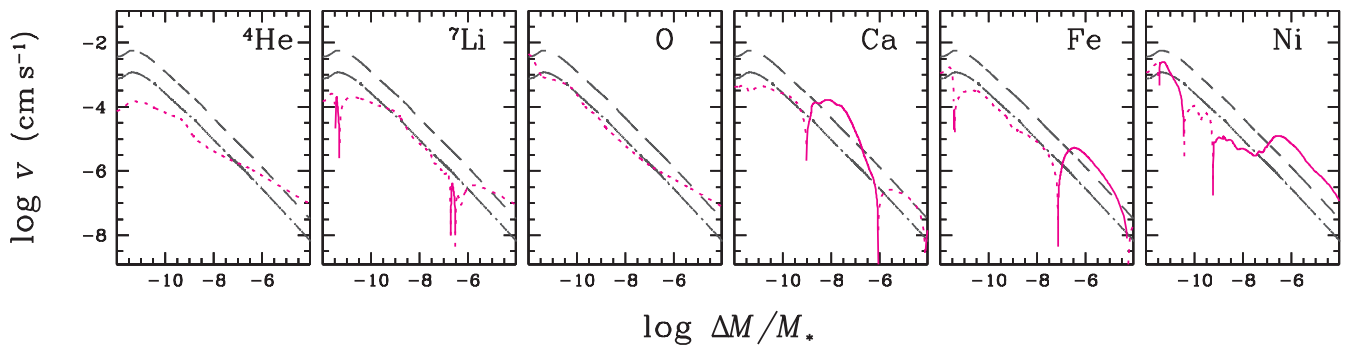}
\caption{Wind velocities (dot-dashed line: $10^{-14}$\Mloss;
dashed line: $5\times 10^{-14}$\Mloss) and diffusion 
velocities (solid when positive, toward the surface, and dotted when 
negative) 
of a few selected elements in a 1.90\Msol{} model near the ZAMS. For most species, wind
velocities decrease more rapidly inwards than diffusion velocities.}\label{fig:vwind}
\end{center}
\end{figure*}
Significant abundance variations appear in the
interior of our PMS models. For example, in the \hbox{2.50W5E-14} model (bottom row of 
Fig.\,\ref{fig:flux}), 
a $1.1$\,dex Ca overabundance develops at $\DM \simeq -9.4$ as early as 4\,Myr. 
It might seem puzzling however that such an 
overabundance transforms into a strong underabundance which reaches below $-$1\,dex at
30\,Myr. Why should the behavior at 4\,Myr be so different from the rest of the star's
evolution and why does a strong overabundance develop in a region 
where \gr(Ca) is below
gravity? The reason stems from the fact that the wind progressively advects matter
from deeper within the star. That depth is simply given by:
\begin{equation}
\Delta M \simeq {\dot M} t.
\label{eqn:flux}
\end{equation}
So, at 4\,Myr, the wind brings to the surface matter which
originates from $4\times 10^{6}$\,yr\,$\cdot\, 5 \times 10^{-14}\Mloss =
2 \times 10^{-7}\Msol \sim 10^{-7}M_*$. 
Correspondingly, 
one sees in the top row
of Fig.\,\ref{fig:flux} that the flux is nearly 
constant from the surface down to that depth
(except over CZs), while it is clearly not 
constant below that depth. Naturally, the depth above 
which the flux is conserved due to advection from the wind increases with age 
(compare the curves at 4, 30 and 300\,Myr). In order to 
conserve the flux, $X$(Ca) at 4\,Myr increases
above $\Dm \sim 10^{-7}$ to compensate 
for the decrease in \gr(Ca)\footnote{The same type of solution was obtained 
for oxygen in \citet{landstreet98}.}. This process can be described by:
\begin{equation}
\mathcal{F}(r)=cst\simeq r^2\rho(U+U_{{\rm w}})c
\label{eqn:cst1}
\end{equation}
where $\mathcal{F}(r)$ is the local flux at radius $r$, $\rho$ is the 
local density, $U$ and $U_{{\rm w}}$ 
are the advective
part of the atomic diffusion velocity and 
wind velocity respectively, and $c$ is concentration\footnote{This equation  
is derived and discussed 
in Sect.\,5.1.1 of Paper I.}. Although gravity is stronger 
than \gr{} over a fraction of the
stellar envelope for all elements shown in Fig.\,\ref{fig:flux}, the downward 
diffusion velocity
is never larger than the wind velocity (see Fig.\,\ref{fig:vwind} for an example with a 1.9\Msol{}
model). 
Therefore, as long as the 
absolute value of the wind velocity is larger than the downward diffusion velocity, any 
given element
is dragged toward the surface, and its local abundance adjusts in order to
conserve flux. Notice that the fraction of the envelope $\Delta M$ which can be 
described by Eq.\,[\,\ref{eqn:cst1}] increases with time since the wind progressively 
advects more mass toward the surface.

With this in mind, one can understand that at $t = 4$\,Myr, 
the increase of $X$(Ca) above $\Dm \sim 10^{-7}$
is generated by the large flux arriving from regions where 
\gr(Ca) is larger than gravity and which is conserved even as \gr(Ca) decreases toward the surface.
As time passes, Ca arrives at the surface from deeper within the star, where \gr(Ca) is much 
smaller than its value at
$\Dm \sim 10^{-7}$, so that the flux is smaller, the flux conservation applies over a larger mass 
and the surface 
abundance consequently decreases.
This shows the importance of a solution over the whole star, as done here, since 
applying an inner boundary condition at 
$\Dm \sim 10^{-7}$ would lead to an erroneous solution after 4\,Myr. 

On the PMS, the internal concentration variations are 
much smaller for $^7$Li, O and Fe than for Ca. However,
after the star arrives on the MS, as illustrated by the curve at 30\,Myr, larger 
variations appear, including a 
nearly $\sim$0.8\,dex overabundance of Fe which spans from the surface down to $\DM \sim -8$. 


As shown in Fig.\,\ref{fig:vwind}, for a mass loss rate 
of $10^{-14}\Mloss$, the downward diffusion velocity 
is greater than the wind velocity for some elements, whereby the flux conservation 
regime as described by Eq.\,[\,\ref{eqn:cst1}] 
cannot be extended to all
elements. The regime shift approximately occurs at
$2 \times 10^{-14}\Mloss$ (see Fig.\,5 of Paper I and corresponding discussion). This 
mass loss rate also marks the limit below which iron 
accumulation near $T\sim 200\,000$\,K can lead 
to iron peak convection
(see Fig.\,5 of Paper I). 

\subsection{Surface abundances}
\label{sec:surface}
An element's surface abundance depends on age, stellar mass and mass loss rate. 
In Fig.\,\ref{fig:HR}, the surface abundances of He, $^7$Li, Ca and Fe 
are shown for models of 1.5, 1.9, 2.5 and 2.8\Msol{}
with a mass loss rate of $5 \times 10^{-14}\Mloss$. 
As stellar mass increases, anomalies appear at the surface
earlier. 
For instance, at around 3\,Myr the 2.80W5E-14 model
has a 0.6\,dex overabundance of Ca while all other models still have their initial 
abundances (Fig.\,\ref{fig:HR}{\it g}). The same can be said 
for Fe overabundances or He underabundances, which 
appear later in smaller stellar mass models. Nonetheless, it 
is evident that for a given element, 
the overall shape
of the surface abundance evolution curve is very similar for the three heavier models. 
This is due to two things: the appearance of a separate He\,II CZ, as well as 
flux conservation as described in the previous section.

For all four elements shown in Fig.\,\ref{fig:HR}, 
the initial, short-lived 
abundance maxima are caused by evolutionary effects. 
The appearance of a radiative zone which separates the He\,II CZ from the SCZ allows for 
chemical separation to occur near the surface. The direction of the anomaly is determined
by \gr{}-$g$ within that region (between $\DM\sim -10.5$ and $ -9$), and thus He, $^7$Li and Ca 
become underabundant,
whereas Fe develops an overabundance (compare middle panel of Fig.\,\ref{fig:flux} with 
Fig.\,\ref{fig:HR}). Following this brief episode, 
the surface abundance is determined by flux conservation, and so results from chemical separation 
occuring deeper within the star\footnote{Thermohaline convection, as 
described in \citet{theado09}, should not affect 
our results as the small mean molecular weight
gradients that materialize occur in or just below the He\,II CZ, and should not affect surface 
abundances --- after a fraction of a Myr (see Sect.\,\ref{sec:structure}) --- if the internal 
solution is dominated by the wind (e.g. $ {\dot M} \gtrsim 2 \times 10^{-14}$, 
see also Fig.\,9
and discussion in Sect.\,8.1 of Paper I).}.
For instance, the subsequent, gradual $X$(Fe)
increase results from the wind slowly advecting matter which originated below $\DM \sim -7$ 
(for the 2.50W5E-14 model), where 
\gr(Fe) is greater than $g$. Similarly, the variations of $X$($^7$Li) which materialize at the surface 
are due to variations seen in \gr($^7$Li) between $\DM \sim -7.5$ and $-6$.  

The amplitude of the anomalies also depends on stellar mass. The Ca overabundance 
that materializes on the PMS in the
three heavier models reaches 0.65\,dex for the \hbox{1.90W5E-14} model, 
while it reaches 1.1\,dex in the 2.50W5E-14
model. The He and $^7$Li 
underabundances, as well as the initial Fe overabundance follow the same dependence. 
The 1.50W5E-14 model behaves
differently since most of the mass interval from $\DM \sim -7.5$ and $-6$ 
is mixed by convection (see Fig.\,\ref{fig:grga}).
 
\subsubsection{Effect of varying the mass loss rate}
\label{massloss}
\begin{figure*}[!ht]
\begin{center}
\includegraphics[scale=0.99]{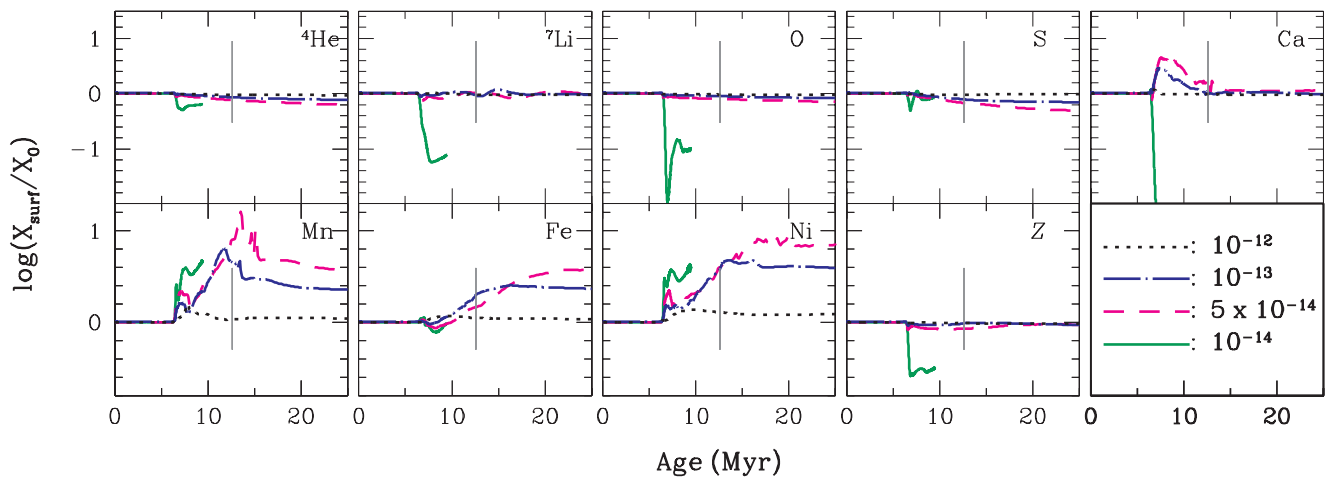}
\caption{Surface abundance variations for models of 1.9\Msol{} 
with different mass loss rates which are identified in the lower right panel in $\Mloss{}$. The vertical line
indicates the approximate end of the PMS.}
\label{fig:absurf1.9}
\end{center}
\end{figure*}
Generally, increasing the mass loss rate lowers the amplitude of surface abundance anomaly extrema, 
as sinking elements
are further advected toward the surface while supported elements are more effectively evacuated. 
In Fig.\,\ref{fig:absurf1.9}, the surface abundances are shown for a 1.9\Msol{} model with
four different mass loss rates. For this model, the PMS ends near 10\,Myr (see Fig.\,\ref{fig:HR}{\it c}). 
For the 1.90W5E-14 and 1.90W1E-13 models, He, O and S decrease slowly and monotonically, 
while Li oscillates around
its initial value. A calcium overabundance occurs at $\sim$7\,Myr, after which $X$(Ca) returns
to its 
original value near the end of the PMS. Iron peak elements are overabundant for both models throughout
the PMS and onto the MS, with Fe maxima of 0.4 and 0.6\,dex for the 1.90W1E-13 
and \hbox{1.90W5E-14} models respectively. For all elements, maximum amplitudes are 
smaller when the mass loss rate is larger.  

The 1.90W1E-14 behaves quite differently: large anomalies appear very rapidly, 
especially for sinking elements; $^7$Li and O underabundances reach $-$1.2 and $-$2\,dex 
respectively, while 
Ca is more than 100 times underbundant as soon as chemical separation reaches the surface. This is 
because, in contrast to the other two mass loss rates, the wind generated by a mass loss 
rate of $10^{-14}\Mloss$ is 
not strong enough to dominate inward diffusion for most sinking elements. In Fig.\,\ref{fig:vwind},
for instance, 
one can see that for this mass loss rate, the inward diffusion velocity of $^7$Li, O and Ca is 
greater or equal to  
the wind velocity over an important fraction of the upper layers. 
On the other hand, supported elements behave as in the other two models, though 
with larger overabundances
(at least up to the age of the last converged model). This is logical 
since the weaker wind cannot evacuate as much of the metal rich material 
which has accumulated in the SCZ.  

On the PMS, and for this stellar mass, only the \hbox{1.90W1E-14} model generates sufficient anomalies to cause $Z$ 
to vary significantly from its original value. The 1.90W1E-12 model only allows very 
small abundance anomalies at the surface ($\sim 0.1\,$dex).

\subsubsection{Comparison to turbulence models}
\label{sec:turbulence}
In the models of \citet{richer00}, turbulent mixing down to $T\sim 200\,000$\,K, the 
depth of the iron peak CZ, is assumed to be the process which competes 
with atomic diffusion in the stellar interior. However, due to efficient mixing, abundance anomalies
do not appear at the surface of these models until they have arrived on the MS. 
The significant 
overabundance of Ca in mass loss models (Fig.\,\ref{fig:HR}{\it f}, see also \citealt{alecian96}), 
is not obtained in the above mentioned models with turbulence, and so could 
be used as an observational test for young stars.
 
%

\section{Comparison to observations}
\label{sec:observations}
Confronting our models to observations can offer constraints on mass loss rates 
as well as on mixing processes
near the surface, such as convection. However, in order to compare models with 
observations, one must 
try to fit various parameters
simultaneously, including age, mass (or \teff) and initial surface composition. 
To reduce the arbitrariness of the comparison, we chose three binary 
systems, since both components of such systems should have the same 
age and initial abundances. Unfortunately, 
determining this age and initial
abundance mix is difficult. For all three systems, the age deduced by observers was determined 
using model isochrones
in which many assumptions were necessary. For instance, in all three 
cases atomic diffusion was neglected, and large mass
loss or accretion rates ($\sim 10^{-8}\Mloss$) were assumed. For this reason, we 
believe that it is preferable to fit 
the position in the HR diagram from this study rather than use the 
age determined by other isochrones. 
As for initial abundances, two of the systems have a chemically ``normal" secondary
which offers a glimpse into the initial metal content, even though ``normal" abundances can mean different things
depending on the solar mix used as a reference\footnote{There is still much debate on 
the newer \citet{asplund05} and \citet{asplund09} initial solar abundance mix for which abundances of 
CNO are significantly lower
than the older solar mix of \citet{grevesse96}; the latter are in better agreement with helioseismology models 
(\citealt{delahaye06}).}. However, assuming that both components of a given binary system 
have the same initial abundances allows us 
to make a differential comparison of surface abundances and conclude that differences between the two are caused by
internal processes such as atomic diffusion. The chosen original abundances aren't crucial 
for our analysis since 
our objective here is to show $when$ abundance anomalies appear on the PMS,
which is not very sensitive to the initial abundances used to construct the models 
(see Fig.\,13 and Sect.\,8.1 of Paper I). 

\begin{figure}[!ht]
\begin{center}
\includegraphics[scale=0.55]{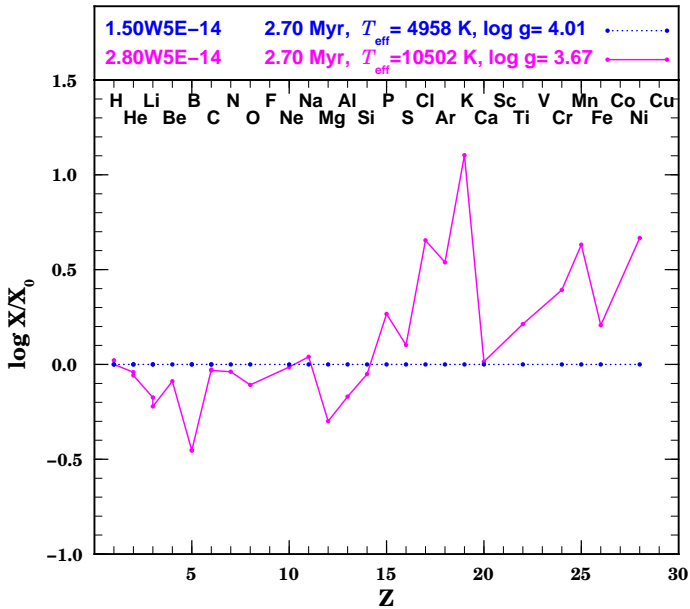}
\caption{Surface abundances for models 
representing the components of the binary star V380 Ori. 
The models were chosen at the age which best
fits the primary in the HR diagram (Fig.\,\ref{fig:HR}{\it a}). In both
models, the mass loss rate is $5\times 10^{-14}\Mloss$.}
\label{fig:V380}
\end{center}
\end{figure}
In Fig.\,\ref{fig:HR}, models are compared to three observed binary systems: 
V380 Ori ($\blacksquare$), HD72106 ($\blacktriangle$) and RS Cha ($\bullet$), which are 
all believed to be HAeBe stars. 
For the Ae star V380 Ori (\citealt{alecian09}), 
the 2.87\Msol{} primary and  
1.6\Msol{} chemically normal secondary are compared to the 2.80W5E-14 and \hbox{1.50W5E-14} 
models respectively. Within the error bars,
the fit in the HR diagram is very good and the corresponding age of the 
primary is $\sim$2.7\,Myr (the age given in \citealt{alecian09}
is 2$\pm 1$\,Myr). The 2.80W5E-14 model has developed overabundances of Fe and Ca which reach 
$\sim 0.3$\,dex and $\sim$0.6\,dex respectively. 
Though this model could only be converged until 2.8\,Myr, it has already developed anomalies which  
explain the high metallicity ([M/H]=0.5) determined in \citet{alecian09}. As illustrated in Fig.\,\ref{fig:V380}, most
metals heavier than $Z \geq 15$ are overabundant by a factor \hbox{of $\sim$2$-$3}, while 
CNO are barely underabundant.
Lithium becomes $-$0.2\,dex underabundant. At the best fitting age for the
primary, the 
1.50W5.14 model 
still has its initial abundances, which also agrees 
with observations. The potential effect of a magnetic field, which is 
observed in the primary, will be discussed in Sect.\,\ref{sec:conclusions}.

For HD\,72106, the 2.4\Msol{} primary and the 1.9\Msol{} secondary are compared to the 2.50W5E-14 and 
1.90W5E-14 models
respectively. In the HR diagram, the fit is not perfect, especially for the primary, though this was also
problematic in Fig.\,4 of \citet{folsom08}. In fact, these authors suggest that at a determined 
age of 6$-$13\,Myr (our best fit model is between 5--7\,Myr), the primary is most likely 
on the ZAMS rather than the PMS. Nonetheless, they found that the
primary was chemically anomalous with important overabundances of iron peak elements 
and a strong underabundace of He,
while the secondary is almost solar. 
In the best fitting models for the primary (5--7\,Myr), iron overabundances vary
from about 0.1 to 0.3\,dex, while Ca overabundances vary from about 1\,dex down to about 0.3\,dex.  For the
same age interval, the He underabundance varies from $-$0.1 to $-$0.15\,dex. These amplitudes are smaller than
the observed values. This could be due to the mass loss rate being too large (see
Fig.\,\ref{fig:absurf1.9}). 
The presence of a magnetic field and of phase
variations (see Fig.\,9 of \citealt{folsom08}) do not justify trying to achieve a better fit. A precise model would require
2 or 3\,D calculations. Furthermore, 
some uncertainty remains on the exact amplitudes since all their abundances were determined simultaneously by fitting 
the observed spectra for a single averaged phase. Over this 
same age interval, the secondary still has its initial abundances, which agrees with observations. 


For RS Cha, the 1.89\Msol{} primary and the 1.87\Msol{} secondary may be compared to the 1.90W5E-14 model. 
In the HR diagram, the slight discrepancy with the model can be explained by the slight difference 
in stellar mass. Within reasonable error bars, the model is either chemically 
normal (up to 6.7\,Myr), or has developed an overabundance of Ca of about 0.4\,dex accompanied by small 
underabundances of $^7$Li and He of $-$0.11\,dex and $-$0.05\,dex respectively. 
This model does not 
seem to explain the iron enrichment factor of 1.5 
obtained by the authors. However, in contrast to the two previous binary 
systems, both these stars have the same composition; therefore,
there is no difference in composition between the stars to explain, and  
so the initial abundances could be responsable
for the abundance anomalies with respect to the solar composition. The X-ray emission 
observations of \citet{mamajek99} suggest that accretion could play a role. A smaller mass
loss rate would also lead to larger anomalies.   

Other young single HAeBe stars which are not shown in Fig.\,\ref{fig:HR} may 
also be compared to our models. The star 
HD\,104237 has a mass of about 2.3\Msol{} 
($\teff=8000$\,K), a luminosity of about 1.42\,L$_{\odot}$ (\citealt{bohm04})
and an approximate age of 
2\,Myr\,(\citealt{vandenAncker98}), as 
well as a
magnetic field of about 50\,G \citep{donati97,donati00}. 
It is an HAeBe 
star for which \citet{acke04} found that Si, Cr and Fe abundances were 
solar, which agrees with our results since 
even the heavier \hbox{2.50W5E-14} model is roughly normal 
until 3\,Myr (see Fig.\,\ref{fig:HR}). 
Similarly, HD\,190073, which has an age of
1.2$\pm$0.6\,Myr and a mass of 2.85$\pm$0.25\Msol{} ($\teff=9250\,$K), was also found to be roughly solar 
(\citealt{acke04}). Within the timescales shown in Fig.\,\ref{fig:HR}, this is also compatible 
with our results. Finally, 
though it has just recently embarked on the MS, the young magnetic Bp cluster star NGC 2244-344 
which has $\teff\simeq 15\,000\,$K and an age of 2\,Myr
(\citealt{bagnulo04}) can also be compared to our results. A rough 
extrapolation suggests that the observed 
$\sim$1\,dex overabundances of Si and Fe and $\sim$2\,dex overabundances of Ti and Cr, as well 
as the 
$\sim$1\,dex underabundance of He could simply be the 
result of chemical separation which began on the PMS. 
 
\section{General discussion and conclusions}
\label{sec:conclusions}

Though it is often assumed negligeable for A and F type PMS stars, 
chemical separation resulting 
from atomic diffusion can affect both the
surface and interior of these young stars. The mass in  
convection zones (Fig.\,\ref{fig:grga}) and 
internal concentrations (Fig.\,\ref{fig:flux}) can be modified \emph{before} stars 
arrive on the MS. The amplitude of the internal
concentration variations depends on stellar mass. Equivalently, the age at which 
abundance anomalies 
appear at the surface also depends on stellar mass. 
In the presence of weak mass loss, and 
for models with no turbulent mixing, 
rotation or magnetic fields, significant internal variations and surface 
anomalies (both exceeding factors of 10 for some elements) appear as 
early as $\sim$2\,Myr in early-type A stars, and \hbox{$\sim$20-25\,Myr} in
cooler F stars. 

Mass loss only slightly affects the age at which abundance 
anomalies occur at the surface, although its 
inclusion in our calculations 
was necessary in order to evolve models to ages of interest. 
This being said, mass loss does affect the maximum amplitude of surface anomalies
(see Fig.\,\ref{fig:absurf1.9}); a mass loss rate $\geq 10^{-12}\Mloss$ 
nearly flattens surface abundance anomalies, 
whereas mass loss rates $\leq 10^{-14}\Mloss$ allow
strong underabundances of elements which are not supported by the radiation field, 
and may allow the appearance of an iron peak CZ. 

The important Ca overabundance materializing in our models of PMS stars, which 
is not
obtained in similar models with turbulence, could help distinguish between these two
scenarios (see Sect.\,\ref{sec:turbulence}). If 
abundance anomalies observed at the surface of A and F type 
PMS stars are in fact due to chemical separation, this 
strongly favors the mass loss model as presented in this paper and should be confirmed
by asteroseismology (see also Sect.\,6 of Paper I). More precise
determinations of individual elemental abundances may also allow further differentiation.

Atomic diffusion coupled with a mass loss rate which is compatible with
observations of AmFm stars may elucidate why there are PMS binary systems 
such as V380\,Ori and HD72106
for which the primary is chemically anomalous while the secondary remains
roughly normal (see Sect.\,\ref{sec:observations}). Since diffusion timescales 
vary more rapidly with stellar mass than with 
mass loss rate (compare Figs.\,\ref{fig:HR} and \ref{fig:absurf1.9}), 
smaller mass loss rates (or atomic diffusion on its own) may also explain these systems. 
Observed abundance anomalies 
in other single PMS stars and young ApBp
stars may also result from atomic diffusion (Sect.\,\ref{sec:observations}), without 
the need for other more exotic
explanations. This being said, many PMS stars 
have observed 
magnetic fields, strong accretion rates (from stellar disks) or both. 
These phenomena 
could also have an effect on transport in the external regions of the star. 

A magnetic field can impact chemical transport 
through its effects on rotation via magnetic braking, by 
modifying convection in the atmosphere (\citealt{cattaneo03}), as well as 
by affecting atomic diffusion velocities (\citealt{alecian06}). In single stars for 
which rotation cannot be slowed 
by tidal forces, strong magnetic fields
may be the only process which sufficiently reduces rotation to allow chemical
separation. Although it has to our knowledge never been shown  
that magnetic fields 
can completely eliminate 
convection, their topologies can induce 
horizontally dependent convection 
and
abundance profiles (\citealt{babel91}), in addition to 
anisotropic mass loss and/or accretion (\citealt{theado05}). If mass loss is in fact anisotropic, 
then strong mass loss rates, as suggested for young PMS stars by 
\citealt{boehm95}, may be present only where field lines are vertical, and thus be compatible with 
anisotropic surface anomalies.
Similarly, magnetic fields may extend the upper limit of $v_{{\rm eq}} = 100\,$km\,s$^{-1}$, which 
was found to eliminate the effects of atomic diffusion (\citealt{charbonneau88}). 
This may allow anomalies to develop even in the fast rotators among HAeBe stars (\citealt{davis83}).

However, until these processes are better 
understood, and magnetic field geometries are better constrained for individual stars, it would
be ill-advised to introduce additional magnetic field related parameters into our calculations.  
The effects of magnetic fields on chemical transport
may best be grasped through comparative mapping of 
superficial magnetic fields and abundances as done for roAp stars in
\citet{kochukhov10} and \citet{luftinger10}, as well as for HAeBe stars in \citet{folsom08}. 


Accretion also affects elemental distribution in the atmospheres of stars. For accretion rates
greater than approximately $10^{-12}\Mloss$, the abundance profiles 
in the atmosphere, and in the interior, simply reflect 
those of the accreted material (\citealt{turcotte93}). For accretion rates a few times larger than  
$10^{-14}\Mloss$, the atmospheric abundances can preserve the accreted material signature while allowing for
abundance gradients due to chemical separation to develop below the SCZ. For even smaller rates, 
atmospheric abundances essentially reflect the result of chemical separation as if there were no accretion. 
In any case, as soon as the star stops
accreting, chemical separation 
resulting from atomic diffusion dominates within 1\,Myr. 

In terms of stellar modelling, HAeBe stars are extremely complex: stellar winds, 
accretion, rotation and magnetic fields complicate simulations. Current models cannot account 
for all these processes without
invoking multiple parameters which may blur any actual 
physics taking place within these stars. Though there are 
other processes involved, it is shown
that atomic diffusion can lead to abundance anomalies on the PMS, and
that neglecting its effects could have an impact on calibrating atmosphere models.

 


\acknowledgements{M.\,Vick would like to thank E.\,Alecian and 
C.\,Folsom for some very insightful
discussions. He also thanks the D\'epartement de physique de 
l'Universit\'e de Montr\'eal for financial support, as well
as everyone at the GRAAL in Montpellier for their amazing hospitality. 
We acknowledge 
the financial support of Programme National de Physique 
Stellaire (PNPS) of CNRS/INSU, France. This research was
partially supported by NSERC at the Universit\'e de Montr\'eal. Finally, we  
thank the R\'eseau qu\'eb\'ecois de calcul de haute performance (RQCHP) for providing us with the
computational resources required for this work.}


\newpage

\begin{thebibliography}{41}
\expandafter\ifx\csname natexlab\endcsname\relax\def\natexlab#1{#1}\fi

\bibitem[{{Acke} \& {Waelkens}(2004)}]{acke04}
{Acke}, B. \& {Waelkens}, C. 2004, \aap, 427, 1009

\bibitem[{{Alecian} {et~al.}(2005){Alecian}, {Catala}, {van't Veer-Menneret},
  {Goupil}, \& {Balona}}]{alecian05}
{Alecian}, E., {Catala}, C., {van't Veer-Menneret}, C., {Goupil}, M., \&
  {Balona}, L. 2005, \aap, 442, 993

\bibitem[{{Alecian} {et~al.}(2008{\natexlab{a}}){Alecian}, {Catala}, {Wade},
  {Donati}, {Petit}, {Landstreet}, {B{\"o}hm}, {Bouret}, {Bagnulo}, {Folsom},
  {Grunhut}, \& {Silvester}}]{alecian08a}
{Alecian}, E., {Catala}, C., {Wade}, G.~A., {et~al.} 2008{\natexlab{a}},
  \mnras, 385, 391

\bibitem[{{Alecian} {et~al.}(2008{\natexlab{b}}){Alecian}, {Wade}, {Catala},
  {Bagnulo}, {Boehm}, {Bohlender}, {Bouret}, {Donati}, {Folsom}, {Grunhut}, \&
  {Landstreet}}]{alecian08b}
{Alecian}, E., {Wade}, G.~A., {Catala}, C., {et~al.} 2008{\natexlab{b}}, \aap,
  481, L99

\bibitem[{{Alecian} {et~al.}(2009){Alecian}, {Wade}, {Catala}, {Bagnulo},
  {B{\"o}hm}, {Bouret}, {Donati}, {Folsom}, {Grunhut}, \&
  {Landstreet}}]{alecian09}
{Alecian}, E., {Wade}, G.~A., {Catala}, C., {et~al.} 2009, \mnras, 400, 354

\bibitem[{{Alecian}(1996)}]{alecian96}
{Alecian}, G. 1996, \aap, 310, 872

\bibitem[{{Alecian} \& {Stift}(2006)}]{alecian06}
{Alecian}, G. \& {Stift}, M.~J. 2006, \aap, 454, 571

\bibitem[{{Asplund} {et~al.}(2005){Asplund}, {Grevesse}, \&
  {Sauval}}]{asplund05}
{Asplund}, M., {Grevesse}, N., \& {Sauval}, A.~J. 2005, in Astronomical Society
  of the Pacific Conference Series, Vol. 336, Cosmic Abundances as Records of
  Stellar Evolution and Nucleosynthesis, ed. T.~G. {Barnes}, III \& F.~N.
  {Bash}, 25

\bibitem[{{Asplund} {et~al.}(2009){Asplund}, {Grevesse}, {Sauval}, \&
  {Scott}}]{asplund09}
{Asplund}, M., {Grevesse}, N., {Sauval}, A.~J., \& {Scott}, P. 2009, \araa, 47,
  481

\bibitem[{{Babel} \& {Michaud}(1991)}]{babel91}
{Babel}, J. \& {Michaud}, G. 1991, \apj, 366, 560

\bibitem[{{Bagnulo} {et~al.}(2004){Bagnulo}, {Hensberge}, {Landstreet},
  {Szeifert}, \& {Wade}}]{bagnulo04}
{Bagnulo}, S., {Hensberge}, H., {Landstreet}, J.~D., {Szeifert}, T., \& {Wade},
  G.~A. 2004, \aap, 416, 1149

\bibitem[{{Boehm} \& {Catala}(1995)}]{boehm95}
{Boehm}, T. \& {Catala}, C. 1995, \aap, 301, 155

\bibitem[{{B{\"o}hm} {et~al.}(2004){B{\"o}hm}, {Catala}, {Balona}, \&
  {Carter}}]{bohm04}
{B{\"o}hm}, T., {Catala}, C., {Balona}, L., \& {Carter}, B. 2004, \aap, 427,
  907

\bibitem[{{Catala} {et~al.}(2007){Catala}, {Alecian}, {Donati}, {Wade},
  {Landstreet}, {B{\"o}hm}, {Bouret}, {Bagnulo}, {Folsom}, \&
  {Silvester}}]{catala07}
{Catala}, C., {Alecian}, E., {Donati}, J., {et~al.} 2007, \aap, 462, 293

\bibitem[{{Cattaneo} {et~al.}(2003){Cattaneo}, {Emonet}, \&
  {Weiss}}]{cattaneo03}
{Cattaneo}, F., {Emonet}, T., \& {Weiss}, N. 2003, \apj, 588, 1183

\bibitem[{{Charbonneau} \& {Michaud}(1988)}]{charbonneau88}
{Charbonneau}, P. \& {Michaud}, G. 1988, ApJ, 327, 809

\bibitem[{{Cox}(1968)}]{cox68}
{Cox}, J.~P. 1968, {Principles of stellar structure - Vol.1: Physical
  principles; Vol.2: Applications to stars} (New York: Gordon and Breach, 1968)

\bibitem[{{Davis} {et~al.}(1983){Davis}, {Strom}, \& {Strom}}]{davis83}
{Davis}, R., {Strom}, K.~M., \& {Strom}, S.~E. 1983, \aj, 88, 1644

\bibitem[{{Delahaye} \& {Pinsonneault}(2006)}]{delahaye06}
{Delahaye}, F. \& {Pinsonneault}, M.~H. 2006, \apj, 649, 529

\bibitem[{{Donati} {et~al.}(2000){Donati}, {Mengel}, {Carter}, {Marsden},
  {Collier Cameron}, \& {Wichmann}}]{donati00}
{Donati}, J., {Mengel}, M., {Carter}, B.~D., {et~al.} 2000, \mnras, 316, 699

\bibitem[{{Donati} {et~al.}(1997){Donati}, {Semel}, {Carter}, {Rees}, \&
  {Collier Cameron}}]{donati97}
{Donati}, J., {Semel}, M., {Carter}, B.~D., {Rees}, D.~E., \& {Collier
  Cameron}, A. 1997, \mnras, 291, 658

\bibitem[{{Folsom} {et~al.}(2008){Folsom}, {Wade}, {Kochukhov}, {Alecian},
  {Catala}, {Bagnulo}, {B{\"o}hm}, {Bouret}, {Donati}, {Grunhut}, {Hanes}, \&
  {Landstreet}}]{folsom08}
{Folsom}, C.~P., {Wade}, G.~A., {Kochukhov}, O., {et~al.} 2008, \mnras, 391,
  901

\bibitem[{{Gonzalez} {et~al.}(1995){Gonzalez}, {LeBlanc}, {Artru}, \&
  {Michaud}}]{gonzalez95}
{Gonzalez}, J.-F., {LeBlanc}, F., {Artru}, M.-C., \& {Michaud}, G. 1995, ApJ,
  297, 223

\bibitem[{{Grevesse} {et~al.}(1996){Grevesse}, {Noels}, \&
  {Sauval}}]{grevesse96}
{Grevesse}, N., {Noels}, A., \& {Sauval}, A.~J. 1996, in Astronomical Society
  of the Pacific Conference Series, Vol.~99, Cosmic Abundances, ed. S.~S.
  {Holt} \& G.~{Sonneborn}, 117

\bibitem[{{Herbig}(1960)}]{herbig60}
{Herbig}, G.~H. 1960, \apjs, 4, 337

\bibitem[{{Iben}(1965)}]{iben65}
{Iben}, Jr., I. 1965, \apj, 141, 993

\bibitem[{{Kochukhov} \& {Wade}(2010)}]{kochukhov10}
{Kochukhov}, O. \& {Wade}, G.~A. 2010, \aap, 513, A13+

\bibitem[{{Landstreet} {et~al.}(1998){Landstreet}, {Dolez}, \&
  {Vauclair}}]{landstreet98}
{Landstreet}, J.~D., {Dolez}, N., \& {Vauclair}, S. 1998, \aap, 333, 977

\bibitem[{{LeBlanc} {et~al.}(2000){LeBlanc}, {Michaud}, \&
  {Richer}}]{leblanc00}
{LeBlanc}, F., {Michaud}, G., \& {Richer}, J. 2000, ApJ, 538, 876

\bibitem[{{L{\"u}ftinger} {et~al.}(2010){L{\"u}ftinger}, {Kochukhov},
  {Ryabchikova}, {Piskunov}, {Weiss}, \& {Ilyin}}]{luftinger10}
{L{\"u}ftinger}, T., {Kochukhov}, O., {Ryabchikova}, T., {et~al.} 2010, \aap,
  509, A71+

\bibitem[{{Mamajek} {et~al.}(1999){Mamajek}, {Lawson}, \&
  {Feigelson}}]{mamajek99}
{Mamajek}, E.~E., {Lawson}, W.~A., \& {Feigelson}, E.~D. 1999, \apjl, 516, L77

\bibitem[{{Richard} {et~al.}(2001){Richard}, {Michaud}, \&
  {Richer}}]{richard01}
{Richard}, O., {Michaud}, G., \& {Richer}, J. 2001, ApJ, 558, 377

\bibitem[{{Richer} {et~al.}(1998){Richer}, {Michaud}, {Rogers}, {Turcotte}, \&
  {Iglesias}}]{richer98}
{Richer}, J., {Michaud}, G., {Rogers}, F., {Turcotte}, S., \& {Iglesias}, C.~A.
  1998, ApJ, 492, 833

\bibitem[{{Richer} {et~al.}(2000){Richer}, {Michaud}, \& {Turcotte}}]{richer00}
{Richer}, J., {Michaud}, G., \& {Turcotte}, S. 2000, ApJ, 529, 338

\bibitem[{{Th{\'e}ado} {et~al.}(2009){Th{\'e}ado}, {Vauclair}, {Alecian}, \&
  {Le Blanc}}]{theado09}
{Th{\'e}ado}, S., {Vauclair}, S., {Alecian}, G., \& {Le Blanc}, F. 2009, \apj,
  704, 1262

\bibitem[{{Th{\'e}ado} {et~al.}(2005){Th{\'e}ado}, {Vauclair}, \&
  {Cunha}}]{theado05}
{Th{\'e}ado}, S., {Vauclair}, S., \& {Cunha}, M.~S. 2005, \aap, 443, 627

\bibitem[{{Turcotte} \& {Charbonneau}(1993)}]{turcotte93}
{Turcotte}, S. \& {Charbonneau}, P. 1993, \apj, 413, 376

\bibitem[{{Turcotte} {et~al.}(1998){Turcotte}, {Richer}, {Michaud}, {Iglesias},
  \& {Rogers}}]{turcotte98soleil}
{Turcotte}, S., {Richer}, J., {Michaud}, G., {Iglesias}, C.~A., \& {Rogers}, F.
  1998, ApJ, 504, 539

\bibitem[{{van den Ancker} {et~al.}(1998){van den Ancker}, {de Winter}, \&
  {Tjin A Djie}}]{vandenAncker98}
{van den Ancker}, M.~E., {de Winter}, D., \& {Tjin A Djie}, H.~R.~E. 1998,
  \aap, 330, 145

\bibitem[{{Vick} {et~al.}(2010){Vick}, {Michaud}, {Richer}, \&
  {Richard}}]{vick10}
{Vick}, M., {Michaud}, G., {Richer}, J., \& {Richard}, O. 2010, \aap, 521, A62+

\bibitem[{{Wade} {et~al.}(2005){Wade}, {Drouin}, {Bagnulo}, {Landstreet},
  {Mason}, {Silvester}, {Alecian}, {B{\"o}hm}, {Bouret}, {Catala}, \&
  {Donati}}]{wade05}
{Wade}, G.~A., {Drouin}, D., {Bagnulo}, S., {et~al.} 2005, \aap, 442, L31

\end{thebibliography}
\clearpage

\end{document}